\documentclass{aa}
\usepackage{xcolor}
\usepackage{latexsym,graphicx,amssymb,verbatim,float,natbib,amsmath}
\usepackage{txfonts}

\begin{document}

\title{Origin of the break in the cosmic-ray electron plus positron spectrum at $\sim\,1$~TeV}
\author{Satyendra Thoudam\thanks{E-mail: satyendra.thoudam@ku.ac.ae}}
\institute{Department of Physics, Khalifa University, PO Box 127788, Abu Dhabi, United Arab Emirates}
\date{\today}

\abstract{
Recent measurements of the cosmic-ray electron plus positron spectrum in several experiments have confirmed the presence of a break at $\sim\,1$~TeV. The origin of the break is still not clearly understood. In this work, we explored different possibilities for the origin, which include an electron source spectrum with a broken power law, a power law with an exponential or super-exponential cutoff, and the absence of potential nearby cosmic-ray sources. Based on the observed electron plus positron data from the DAMPE and the H.E.S.S experiments, and considering supernova remnants as the main sources of cosmic rays in the Galaxy, we find statistical evidence in favor of the scenario with a broken power-law source spectrum, with the best-fit source parameters obtained as $\Gamma=2.39$ for the source spectral index, $E_0\approx 1.6$~TeV for the break energy, and $f=1.59\times 10^{48}$~ergs for the amount of supernova kinetic energy injected into cosmic-ray electrons. This power-law break in the spectrum has been predicted for electrons confined inside supernova remnants after acceleration via diffusive shock acceleration process, and also indicated by the multi-wavelength study of supernova remnants. All of this evidence shows that the observed spectral break provides a strong indication of a direct link between cosmic-ray electrons and their sources. Our findings further show that electrons must undergo spectral changes while escaping the source region in order to reconcile the difference between the spectral index of electrons observed inside supernova remnants and that obtained from Galactic cosmic-ray propagation studies.
}

\authorrunning{Thoudam}
\titlerunning{Origin of the break in the cosmic-ray electron spectrum at $\sim\,1$~TeV}
\maketitle

\section{Introduction}
\label{sec-Intro}
High-energy cosmic-ray (CR) electrons and positrons with energies above $\sim\,10$~GeV suffer radiative losses mainly through synchrotron and inverse Compton processes during their propagation through the Galaxy. For a power-law source spectrum of $E^{-\Gamma}$ as predicted by the diffusive shock acceleration (DSA) theory (e.g., \citealp{Drury1983}), radiative losses coupled with an energy-dependent diffusive propagation of CRs are expected to produce an equilibrium electron spectrum in the Galaxy that follows a power law $E^{-(\Gamma+1-\xi)}$ for stationary sources uniformly distributed throughout the Galactic disk \citep{Ginzburg1976, Atoyan1995}, where $\xi\sim(0.2-0.35)$, depending on the index of the CR diffusion coefficient $D(E)\propto E^a$, which is typically in the range of $a\sim (0.3-0.6)$ in common models of CR propagation in the Galaxy \citep{Jones2001, Strong2010}.

In recent years, unprecedented high-precision measurements of CR electrons plus positrons in several experiments, such as those with the Alpha
Magnetic Spectrometer (AMS-02; \citealp{AMS2019a}), Calorimetric Electron Telescope (CALET; \citealp{CALET2017}), Dark Matter Particle Explorer (DAMPE; \citealp{DAMPE2017}), \textit{Fermi} Large Area Telescope (\textit{Fermi}-LAT; \citealp{Fermi2017}), and the High Energy Stereoscopic System (H.E.S.S.; \citealp{HESS2008, HESS2009}), have revealed spectral structures that deviate from theoretical expectations. Two prominent features are a gradual hardening above $\sim\,(30-50)$~GeV and a sharp break or steepening at $\sim\,1$~TeV ($=10^3$~GeV). Evidence for the break was first reported by the H.E.S.S. experiment and later confirmed by the DAMPE experiment with its high-resolution measurement covering an energy range of $\mathrm{25~GeV-4.6~TeV}$. The observed spectrum closely follows $E^{-3}$ and $E^{-4}$ below and above the break, respectively.

The origin of the spectral break is still not clearly understood. However, it can be understood that the break cannot be due to the observed suppression in the positron flux at high energies, as positrons represent only a small fraction of the combined electron and positron flux. The positron fraction increases steadily from about $5\%$ at $\sim\,8$~GeV to a maximum of about $15\%$ at $\sim\,(300-400)$~GeV, above which it shows a continuous decrease \citep{AMS2019b}. At $\sim\,1$~TeV, positrons make up only about $10\%$ of the combined electron plus positron flux. This indicates that the spectral break has to do with the electron component, which is therefore our main focus in this work.

The break is difficult to explain as an effect of the propagation of CR electrons in the Galaxy. In the standard model of CR propagation in the Galaxy, where we sought a steady-state solution, CR electrons are expected to have a spectrum that steepens continuously with energy without any break as discussed above, unless certain special conditions are imposed \citep{Lipari2019}. Features may arise that are related to the sudden drop in the energy loss rate of high-energy electrons as the inverse Compton scattering cross-section changes from the Thomson limit to the Klein-Nishina regime \citep{Schlickeiser2010}. However, this drop is expected to produce an upturn in the spectrum \citep{Evoli2020}, not a steepening, which is what is shown by the measurements.

The observed break is most likely caused by an effect related to the source properties of CRs, such as the absence of nearby sources and a cutoff or break in the source spectrum itself. At energies where radiative losses are significant, electrons cannot travel far distances in the Galaxy, and the space-time volume that contains the sources contributing to the observed CR electrons becomes smaller (see e.g., \citealp{Lipari2019}). This can lead to only a small number of sources contributing at high energies, and can thus produce a spectrum at the Earth that is exponentially suppressed \citep{Kobayashi2004}. For a typical interstellar medium (ISM) environment and standard CR propagation parameters, electrons of 1~TeV have an average diffusion length of $\sim\,1$~kpc in the Galaxy (see Sect. \ref{sec-prop} for details). Therefore, if potential CR sources are missing or if they are significantly less numerous in the local ISM within $\sim\,1$~kpc of Earth, the observed CR electron flux can be significantly suppressed at TeV energies. Hereafter, we refer to this as the ``missing sources" scenario.

Based on theoretical grounds and observational evidence, supernova remnants (SNRs) are considered to be the most plausible sources of CRs in the Galaxy (e.g., \citealp{Ginzburg1961, Drury1983, Reynolds1999}). Theoretically, the DSA process at the supernova shocks can accelerate supra-thermal particles of the ISM to very high energies. For electrons, the maximum energy can be limited by different factors, which can lead to different shapes of the spectral cutoff at the highest energies. For instance, for the widely adopted case of Bohm-type diffusion of particles, the electron spectrum follows the most commonly adopted form, the "exponential cutoff" $\exp(-E/E_0)$, if the maximum energy is limited by the presence of a free-escape boundary ahead of the shock in the upstream region (see \citealp{Yamazaki2014} and references therein). This type of spectral cutoff is widely used in CR propagation studies, for both the leptonic and hadronic species \citep{Kobayashi2004, Thoudam2016}, as well as in the study of radio and X-ray observations from SNRs \citep{Reynolds1999}. If the maximum energy is limited by radiative losses or by the finite age of the SNR, a ``super-exponential cutoff" $\exp(-E^2/E_0^2)$ is expected \citep{Kang2009, Kang2011}. This shape of the spectral cutoff is primarily expected for cases where the CR diffusion coefficient around the shock has a strong energy dependence, such as in the case of a Bohm-like diffusion \citep{Zirakashvili2007, Blasi2010}. The super-exponential cutoff is also found to explain better the observed synchrotron spectrum from shell-type SNRs when a more correct calculation of synchrotron emission is implemented, instead of the widely adopted $\delta$-function approximation \citep{Zirakashvili2007}. Recently, this form for the spectral cutoff was used to study spectral features in the CR electron spectrum \citep{Fang2018, Evoli2020}. 

Another spectral feature that is naturally expected to be found for electrons present inside SNRs is a break in the power-law spectrum (hereafter, the "broken power-law" scenario). In the DSA process applied to supernova shocks, the majority of the particles escape downstream of the shock after acceleration and remain confined, during which time high-energy electrons suffer radiative losses. At a given age of the remnant, $t_\mathrm{age}$, electrons below a certain energy $E_\mathrm{b}$, whose energy loss time $t_\mathrm{loss}>t_\mathrm{age}$ will maintain the spectrum generated at the shock $E^{-\Gamma}$, while those above $E_\mathrm{b}$ with $t_\mathrm{loss}<t_\mathrm{age}$ will follow a steeper spectrum $E^{-(\Gamma+1)}$ \citep{Thoudam2011, Ohira2017}. Detailed simulations of time-dependent DSA have confirmed the presence of such a break in the power-law electron spectrum inside SNRs \citep{Kang2011}. They found the break at around $1$~TeV when the SNR age is about $1000$~yr, while the exponential cutoff remained at $\sim\,(40-50)$~TeV for a reasonable set of ISM parameters and a downstream magnetic field of $\sim\,100\,\mu$G for the remnant. Recently, \cite{Morlino2021} have shown that generation of this sharp break in the electron spectrum requires a large amplification of the magnetic field in the downstream region of the shock, possibly generated by the combined effect of both CR-related and turbulent instabilities. Also discussed in recent works \citep{Diesing2019, Cristofari2021} is the importance of this efficient magnetic field amplification operating until the late stages of the remnant's evolution, in order to produce a steeper source index of electrons compared to that of protons, as revealed by Galactic CR propagation studies. 

In this work, we compared the different scenarios described above -- missing sources, exponential cutoff, super-exponential cutoff, and broken power-law -- in terms of the quality of fit to the observed CR electron spectrum, and show that the broken power-law scenario best explains the observed break as well as the overall spectrum.

This paper is organized as follows. In Sect. \ref{sec-prop}, we present the CR propagation model used in our study, and in Sect. \ref{sec-scenarios}, the different scenarios for the spectral break are described in detail. Section \ref{sec-data} presents the electron plus positron data that we used to test the different scenarios, and Sect. \ref{sec-fit} describes the spectrum calculation for the different scenarios and the fitting procedure that we adopted for this work. Section \ref{sec-result} presents the fit results, and Sect. \ref{sec-discuss} discusses the results in detail.

\section{Propagation of cosmic-ray electrons in the Galaxy}
\label{sec-prop}
For a stationary and uniform distribution of CR sources in the Galactic disk, the diffusive propagation of CR electrons in the Galaxy subject to synchrotron and inverse Compton radiative losses is described as
\begin{equation}
\label{eq-diffu}
\nabla\cdot(D\nabla N)+\frac{\partial}{\partial E}\lbrace b(E)N\rbrace=-q,
\end{equation}
where $E$ is the kinetic energy, $N(\textbf{r},E)$ is the differential number density at a position $\textbf{r}\equiv(r,\phi,z)$ in cylindrical coordinates (with $r$ representing the radial position, $\phi$ the azimuthal, and $z$ the perpendicular position to the Galactic plane), and $q(\textbf{r},E)$ is the source term denoting the electron injection rate per unit volume into the ISM. The diffusion coefficient is taken as $D(E)=D_0\beta(E/E_0)^a$, where $D_0=1.55\times10^{28}$~cm$^2$~s$^{-1}$ is the diffusion constant, $\beta=v/c$ is the ratio of the particle velocity $v$ to the velocity of light $c$, $E_0=3$~GeV, and $a=0.54$ is the diffusion index \citep{Thoudam2013}. This form of $D(E)$ neglects the break at $\mathrm{\sim\,300~GeV}$ reported in some recent studies based on the secondary-to-primary ratios measured by the AMS-02 experiment \citep{Ferronato2024}. $b(E)$ is the radiative energy loss rate, which is the sum of synchrotron and inverse Compton losses. Following \cite{Schlickeiser2010}, we took
\begin{equation}
\label{eq-b}
b(E)=\frac{4}{3}\sigma_\mathrm{T}c\left(\frac{E}{m_\mathrm{e}c^2}\right)^2\left[U_\mathrm{B}+\sum_{i=1}^4\frac{W_\mathrm{i}E_\mathrm{i}^2}{E^2+E_\mathrm{i}^2}\right],
\end{equation}
where $\sigma_\mathrm{T}$ is the Thomson cross-section, $m_\mathrm{e}c^2$ is the electron rest-mass energy, and $U_\mathrm{B}=B^2/8\pi$ is the energy density of the magnetic field of strength $B$. We took $B=3\,\mu$G. This value of the magnetic field is the average over the $z=\pm 5$~kpc region below and above the Galactic plane, calculated at the galactocentric radius of $8.5$~kpc using the relation $B(z)=B_0\,e^{-|z|/3\mathrm{kpc}}\,\mu\mathrm{G}$ adopted in the GALPROP CR propagation code \citep{Strong2010}. We took $B_0=6\,\mu\mathrm{G}$, the local magnetic field value \citep{Beck2000}. A recent detailed study reports a slightly lower value of the Galactic magnetic field \citep{Unger2024}. The summation in Eq. \ref{eq-b} represents the sum of the inverse Compton losses in the Klein-Nishina limit over four different diffuse photon fields present in the ISM -- (1) photons from spectral type B stars, (2) photons from spectral type G-K stars, (3) infrared radiation, and (4) cosmic microwave background, where $W_\mathrm{i}$ denotes the respective photon energy density, and $E_\mathrm{i}$ is the critical Klein-Nishina energy above which the Thomson cross-section starts to break down. $(W_\mathrm{i}, E_\mathrm{i})$ for the four photon fields are taken as $\mathrm{(0.09~eV/cm^3, 40~GeV)}$, $\mathrm{(0.3~eV/cm^3, 161~GeV)}$, $\mathrm{(0.4~eV/cm^3, 4.0\times10^4~GeV)}$, and $\mathrm{(0.25~eV/cm^3, 3.0\times 10^5~GeV)}$, respectively \citep{Schlickeiser2010}. More details on the relativistic treatment of inverse Compton losses of high-energy electrons can be found in \cite{Delahaye2010}.

As high-energy electrons cannot travel far distances in the Galaxy because of radiative losses, the effect of diffusion boundary of the Galaxy on the solution of Eq. (\ref{eq-diffu}) can be neglected for energies within our scope of interest, that is, above $\sim\,10$~GeV. The Green's function of Eq. (\ref{eq-diffu}) can be obtained as

\begin{eqnarray}
G(\textbf{r},\textbf{r}^\prime,E,E^\prime)=\frac{1}{8\pi^{3/2}b(E)A^{3/2}}  \exp\left(-\frac{(\textbf{r}-\textbf{r}^\prime)^2}{4A}\right)\\
\label{eq-A}
\mathrm{where}\,A(E,E^\prime)=\int_E^{E^\prime}\frac{D(u)}{b(u)}du.
\end{eqnarray}
The CR density at a position $\textbf{r}$ is obtained by convolving the Green's function with the source term $q(\textbf{r},E)$ as
 \begin{equation}
N(\textbf{r},E)=\int^{\infty}_E dE^\prime\int^{\infty}_{-\infty}d\textbf{r}^\prime G(\textbf{r},\textbf{r}^\prime,E,E^\prime) q(\textbf{r}^\prime,E^\prime).
\end{equation}
For a uniform source distribution in the Galactic disk, and assuming azimuthal symmetry, the source term can be written independent of $r$ and $\phi$ as $q(\textbf{r},E)=\nu Q(E)\delta(z)$, where $\nu$ denotes the frequency of supernova explosions per unit surface area in the Galactic disk, and $Q(E)$ is the source spectrum. We took $\nu=25$~Myr$^{-1}$~kpc$^{-2}$, which corresponds to $\sim\,3$ supernova explosions per century in the Galaxy \citep{Grenier2000}. Then, the CR density at the position of the Earth, taken to be $\textbf{r}=0$, due to all the sources being located beyond a radial distance $r_\mathrm{0}$ in the Galactic disk, is obtained as
\begin{align}
\label{eq-sol1}
N(E)&=2\pi\nu\int^{\infty}_E dE^\prime\int^{\infty}_{r_0}r^\prime dr^\prime\int^{\infty}_{-\infty} dz^\prime G(r^\prime,z^\prime,E,E^\prime) Q(E^\prime)\delta(z^\prime)\nonumber\\
&=\frac{\nu}{2\sqrt{\pi}b(E)}\int_E^\infty \frac{Q(E^{\prime})}{\sqrt{A}} \exp\left(-\frac{r_0^2}{4A}\right) dE^{\prime}.
\end{align}
Defining $r_\mathrm{d}(E,E^\prime)=2\sqrt{A(E,E^\prime)}$, Eq. (\ref{eq-sol1}) can be written as
\begin{eqnarray}
\label{eq-sol2}
N(E)=\frac{\nu}{\sqrt{\pi}b(E)}\int_E^\infty \frac{Q(E^{\prime})}{r_\mathrm{d}} \exp\left(-\frac{r_0^2}{r^2_\mathrm{d}}\right) dE^{\prime}.
\end{eqnarray}
$r_\mathrm{d}(E,E^\prime)$ corresponds to the average diffusion length of electron of energy $E^\prime$ before it cools down to energy $E$. For the diffusion coefficient $D(E)\propto E^a$ and energy loss coefficient $b(E)\propto E^2$, as expected in the case of  synchrotron losses and inverse Compton losses in the Thomson regime, we can write an exact analytical expression of Eq. (\ref{eq-A}) as
\begin{eqnarray}
\label{eq-A2}
A(E,E^\prime)= \frac{1}{1-a}\left[D(E)t_\mathrm{loss}(E)-D(E^\prime)t_\mathrm{loss}(E^\prime)\right],
\end{eqnarray}
where $t_\mathrm{loss}(E)\propto1/E$ is the energy loss time for electrons of energy $E$. For the values of the diffusion parameters and the Galactic magnetic field strength of $B=3\,\mu$G used in the present study, the diffusion length for an electron of 1~TeV before losing half its energy was calculated to be $r_\mathrm{d}\sim 1.16$~kpc. In general, for $a<1$ and $E^\prime>>E$, that is, when the electron has lost a significant fraction of its energy ($E=\eta E^\prime$, where $\eta<<1$), from Eq. (\ref{eq-A2}), we have $r_\mathrm{d}\propto\sqrt{D(E)t_\mathrm{loss}(E)}\propto E^{(a-1)/2}$. For $a=0.54$, which is the value adopted in this work, $r_\mathrm{d}$ was found to scale with energy as $r_\mathrm{d}\propto E^{-0.23}$, which shows that high-energy electrons traverse a smaller region of the Galaxy compared to the low-energy ones. This can lead to a significant suppression of the observed electron flux at high energies if potential CR sources are absent in the nearby region, due to the dominant effect of the exponential term in Eq. (\ref{eq-sol2}).

By setting $r_0=0$ in Eq. (\ref{eq-sol2}), the CR electron flux at Earth from a continuous distribution of sources throughout the Galactic disk can be obtained as
\begin{eqnarray}
\label{eq-sol3}
N(E)=\frac{\nu}{\sqrt{\pi}b(E)}\int_E^\infty \frac{Q(E^{\prime})}{r_\mathrm{d}} dE^{\prime}.
\end{eqnarray}
For a single power-law source spectrum in the form $Q(E)\propto E^{-\Gamma}$, Eq. (\ref{eq-sol3}) gives a spectrum at Earth that follows $N(E)\propto E^{-\Gamma-1+\xi}$ for the case of $D(E)\propto E^a$ and $b(E)\propto E^2$. The spectrum steepens mainly due to radiative losses, but with a correction factor $\xi=(1-a)/2$ arising from the diffusive nature of the propagation. For $a=0.54$, the correction factor is $\xi=0.23$.

\section{Different scenarios for the spectral break}
\label{sec-scenarios}
As mentioned briefly in Sect. \ref{sec-Intro}, the observed spectral break at $\sim 1$~TeV is most likely due to an effect related to the sources, such as missing nearby CR sources or an intrinsic cutoff (or break) in the CR source spectrum. In this section, we discuss the different possibilities in detail.

\subsection{Missing sources}
The absence of potential CR sources in the local ISM can lead to a suppression in the CR electron flux at high energies, which may appear as a break or exponential cutoff in the observed spectrum \citep{Kobayashi2004, Lipari2019}. The energy above which the suppression occurs depends on the distance of the closest source. To produce a suppression above $\sim 1$~TeV, the nearest source must be located around $1$~kpc from Earth, as discussed in Sect. \ref{sec-prop} based on Eq. (\ref{eq-sol2}).

The shape of the spectral break in this case is determined by the $\exp\left(-r_0^2/r^2_\mathrm{d}\right)$ term in Eq. (\ref{eq-sol2}). Defining $r_0$ as the diffusion length of electrons corresponding to the break energy $E_0$, we can write $r_\mathrm{0}=2\sqrt{D(E_0)t_\mathrm{loss}(E_0)}\propto E_0^{(a-1)/2}$. For $a=0.54$, this gave $r_0\propto E_0^{-0.23}$, and together with the $r_\mathrm{d}\propto E^{-0.23}$ obtained in Sect. \ref{sec-prop}, the exponential cutoff in Eq. (\ref{eq-sol2}) could be written in terms of energy as $\exp\left[-\left(E/E_0\right)^{0.46}\right]$, which is close to $\exp\left(-\sqrt{E/E_0}\right)$. Therefore, the missing sources scenario is expected to exhibit a spectral cutoff that is harder than the commonly adopted exponential and super-exponential cutoffs, which are discussed below.

\subsection{Exponential cutoff}
In the DSA process, if the maximum energy of the particles is limited by free escape from the acceleration region, the spectral cutoff can follow an exponential cutoff of the form $\exp\left(-E/E_0\right)$ (see e.g., \citealp{Caprioli2009, Ohira2010} and references therein). This type of cutoff is also most commonly used in CR propagation studies (e.g., \citealp{Kobayashi2004, Thoudam2016}), as well as in the study of non-thermal radio and X-ray emission from SNRs \citep{Reynolds1999, Bamba2003}.

In the test particle approximation, assuming the presence of an escape boundary at a distance $L$ ahead of the shock in the upstream region, the maximum energy $E_\mathrm{m}$ of the particles that remain confined is given as the condition $D_\mathrm{s}(E_\mathrm{m})/u=L$, where $D_\mathrm{s}\propto E^\beta$ is the diffusion coefficient of the particles in the accelerator region, and $u$ is the velocity of the shock expanding into a stationary medium. The spectrum at the shock in this case follows \citep{Caprioli2009, Ohira2010, Yamazaki2014}, as
\begin{eqnarray}
\label{eq-sol4}
f_0(E)\propto \exp\left\lbrace -q\int^E \frac{1}{1-\exp\left[-\frac{uL}{D_\mathrm{s}(E^\prime)}\right]}d\log E^\prime\right\rbrace,
\end{eqnarray}
where $q=3r/(r-1)$, with the shock compression ratio $r=4$ for strong shocks. At energies $E\gg E_\mathrm{m}$, the spectrum reduces to $f_0(E)\propto \exp\left[-q\left(E/E_\mathrm{m}\right)^\beta\right]$. For particles diffusing in the Bohm regime, $\beta=1$, which gives a cutoff shape of the form $\exp\left(-E/E_0\right)$, where we took $E_0=E_\mathrm{m}/q$.

\subsection{Super-exponential cutoff}
If the maximum energy of the electrons in the DSA process is limited by radiative losses or by the finite age of the accelerator, the high-energy cutoff is expected to follow a super-exponential cutoff $\exp\left[-\left(E/E_0\right)^2\right]$ for the widely adopted Bohm-type diffusion \citep{Zirakashvili2007, Kang2009, Blasi2010}.

The asymptotic form of the electron spectrum at the shock front in the very high energy regime where the radiative energy losses dominate over acceleration is given as \citep{Zirakashvili2007}
\begin{eqnarray}
\label{eq-sol5}
f_0(E)\propto K_0(E) \exp\left[-S_0(E)\right],
\end{eqnarray}
where, for the same energy dependence of the radiative losses and the diffusion coefficient in the downstream and upstream regions of the shock, $K_0(E)\propto\sqrt{b(E)/E}$, and $S_0(E)$ can be approximated as (see also \citealp{Yamazaki2014})
\begin{eqnarray}
\label{eq-sol6}
S_0(E)\approx \left(\frac{q}{u}\right)^2 \int^E\frac{b(E^\prime)D(E^\prime)}{{E^\prime}^2}dE^\prime.
\end{eqnarray}
Using the condition that the maximum energy is determined by equating the acceleration time $t_\mathrm{acc}(E_\mathrm{m})\approx 4D_\mathrm{s}(E_\mathrm{m})/u^2$ \citep{Bell2013}, with the radiative energy loss time $t_\mathrm{loss}(E_\mathrm{m})\propto1/E_\mathrm{m}$, Eq. (\ref{eq-sol6}) becomes $S_0(E)\approx q\left(E/E_\mathrm{m}\right)^{\beta+1}$. For Bohm-like diffusion, the shape of the spectral cutoff then follows $\exp\left[-\left(E/E_0\right)^2\right]$, where $E_0=E_\mathrm{m}/\sqrt{q}$.

\subsection{Broken power-law}
\label{bpl}
In the standard DSA theory as applied to supernova shock waves, a major fraction of the particles escape downstream of the shock and remain confined within the remnant without crossing the shock again. During the confinement, if high-energy electrons suffer severe radiative losses, this can lead to a power-law break in the cumulative electron spectrum \citep{Thoudam2011, Kang2011, Ohira2017}. However, the presence of such a break requires a large magnetic field amplification in the downstream that continues until the late stages of the SNR evolution \citep{Morlino2021}. This strong field is also required to produce the difference in the source spectral index between protons and electrons above $\sim 10$~GeV that is required in CR propagation studies \citep{Diesing2019, Cristofari2021}.

The temporal evolution of the electron spectrum during the confinement downstream, subject to severe radiative losses and continuous injection into the downstream after acceleration at the shock fronts, can be written as \citep{Kardashev1962, Thoudam2011}
\begin{align}
\label{eq-sol7}
f(E,t)&\propto E^{-(\Gamma+1)}\left[1-\left(1-E/E_0\right)^{\Gamma-1}\right], \; \mathrm{for}\;E<E_0\nonumber \\
&\propto E^{-(\Gamma+1)}, \;\mathrm{for}\; E\geq E_0,
\end{align}
where $\Gamma$ is the spectral index of electrons accelerated at the supernova shocks. Equation (\ref{eq-sol7}) shows that the electron spectrum exhibits a break at an energy, $E_0\propto 1/t$, whose value depends on the age of the remnant, $t$. It can be easily checked using Eq. (\ref{eq-sol7}) that the spectrum below the break follows $E^{-\Gamma}$, while above, the spectrum becomes steeper as $E^{-(\Gamma +1)}$. Interestingly, the electron spectrum measured by the H.E.S.S and the DAMPE experiments also exhibit a similar spectral difference below and above the break energy of $1$~TeV.

\section{Measured electron plus positron spectrum}
\label{sec-data}
Recently, various experiments, such as the AMS-02 \citep{AMS2019a}, CALET \citep{CALET2017}, DAMPE \citep{DAMPE2017}, \textit{Fermi}-LAT \citep{Fermi2017} and H.E.S.S. \citep{HESS2008, HESS2009} instruments, have provided high-quality data on the total electron plus positron spectrum up to energies of a few teraelectronvolts. Their combined spectrum indicates a hardening above $\sim\, (30-50)$~GeV with a power-law index of $\sim\,3.0$ and a steepening above $\sim\,1$~TeV to an index close to $\sim\,4.0$. While the results from the different experiments show a similar behavior over a wide energy range, there are systematic differences between the measured fluxes. For this reason, and also because DAMPE covers an energy range $(\mathrm{25~GeV-4.6~TeV})$ wide enough to detect the presence of the spectral break at $\sim\,1$~TeV in a single measurement, we use the DAMPE data to test the different scenarios for the spectral break described in Sect. \ref{sec-scenarios} together with the H.E.S.S data to increase the statistic at TeV energies.

\begin{table*}
\centering
\caption{\label{table-parameters} Model parameters $(f,\Gamma,E_0, r_0)$ and their ranges considered in the fitting procedure.}
\label{fit-parameter-range}
\resizebox{\textwidth}{!}
{
\begin{tabular}{ccccc|cccc}
\hline
\hline
&\multicolumn{4}{c}{Fit range: $(20-3500)$~GeV}  & \multicolumn{4}{c}{Fit range: $(100-3500)$~GeV}\\
\cline{2-5}\cline{6-9}
&&&&&&&&\\
&$f$                                    &$\Gamma$       &$E_0$          &$r_0$  &$f$                                    &$\Gamma$       &$E_0$  & $r_0$\\
&($\times 10^{49}$~ergs)        &                       &(GeV)          &(kpc)  &($\times 10^{49}$~ergs)  &                       &(GeV)  & (kpc)\\
    
\hline
&&&&&&&&\\
Missing sources         &       $0.1-0.5$       &$2.0-2.3$      & $-$   & $0.5-2.5$       & $0.20-0.60$   &       $1.90-2.25$ & $-$               & $0.5-2.5$\\
Exponential cutoff             &       $0.1-0.3$       &$2.2-2.5$      & $4000-8000$ & $-$       & $0.08-0.30$   &       $2.15-2.45$ & $3000-6000$         & $-$ \\
Super-exponential cutoff       &       $0.1-0.3$       &$2.2-2.5$      & $3000-6000$ & $-$       & $0.08-0.30$   &       $2.15-2.45$ & $2000-5000$         & $-$ \\
Broken power-law                &       $0.1-0.3$       &$2.2-2.5$      & $1000-2000$ & $-$       & $0.08-0.30$   &       $2.15-2.45$ & $1000-2000$         & $-$ \\
\hline
\vspace{0.2cm}
\end{tabular}
}
\tablefoot{The step sizes used for scanning the parameters are: $\Delta f=0.001$~($\times 10^{49}$~ergs), $\Delta\Gamma=0.01$, $\Delta E_0=2$~ GeV and $\Delta r_0=0.05$~kpc.}
\end{table*}

\section{Calculation of CR electron spectrum and fitting the electron plus positron data}
\label{sec-fit}
CR electron spectrum for the different scenarios can be calculated from Eq. (\ref{eq-sol2}) or Eq. (\ref{eq-sol3}) by taking their appropriate source spectrum $Q(E)$. For the missing sources scenario, Eq. (\ref{eq-sol2}) was used by keeping $r_0$ a model parameter, while for the rest, Eq. (\ref{eq-sol3}) was used where $r_0$ was set to $0$. Below are the different forms of $Q(E)$ adopted in our calculation based on the description of the different scenarios given in Sect. \ref{sec-scenarios}:
\begin{align}
\label{eq-source}
Q(E)&=kE^{-\Gamma},\,\mathrm{Missing\,sources}\nonumber\\
&=kE^{-\Gamma}\exp\left(-\frac{E}{E_0}\right),\,\mathrm{Exponential\,cutoff}\nonumber\\
&=kE^{-\Gamma}\exp\left(-\frac{E^2}{E_0^2}\right),\,\mathrm{Super\,exponential\,cutoff}\nonumber\\
&=kE^{-\Gamma}\left[1+\left(\frac{E}{E_0}\right)^{1/\delta}\right]^{-\delta},\,\mathrm{Broken\,power\,law},
\end{align}
where $k$ is a normalization constant that is proportional to the fraction $f$ of the supernova kinetic energy of $10^{51}$~ergs injected into CR electrons, $\Gamma$ is the source index, and $E_0$ is the exponential or super-exponential cutoff. For the broken power-law scenario, the function reproduces the spectral shapes $E^{-\Gamma}$ below the break energy $E_0$ and $E^{-(\Gamma+1)}$ above the break as described by Eq. (\ref{eq-sol7}). The smoothness parameter $\delta$ was fixed at $0.05$ so as to produce a similar level of sharpness at $E_0$ as the spectrum given as Eq. (\ref{eq-sol7}).

For each scenario, the fit to the observed data was performed using three parameters related to the source properties: $(f,\Gamma,r_0)$ for the missing sources and $(f,\Gamma,E_0)$ for the all other cases. We used the fraction $f$ instead of $k$ for the fit as it directly represents the amount of supernova kinetic energy that is converted into CR electrons. In this study, we calculated $f$ on the basis of energy injected into electrons above $1$~GeV. The fit was performed in two steps. First, as $f$ only scales the normalization and does not affect the shape of the final spectrum, we treated it as a nuisance parameter and marginalized it to determine the set of values for $(\Gamma,r_0)$ or $(\Gamma,E_0)$ that gives the maximum likelihood, $\mathcal{L}=\sum_f\mathcal{L}_f$, fit to the observed data. The summation over $f $ was carried out in steps of $\Delta f=0.02$ (in units of $10^{49}$~ergs), and $\mathcal{L}_f=\exp\left({-\chi^2/2}\right)$ with
\begin{equation}
\label{eq-chi2}
\chi^2=\sum_{i}\frac{\left[I_\mathrm{D}(E_\mathrm{i})-I_\mathrm{M}(E_\mathrm{i})\right]^2}{\sigma_\mathrm{D}^2(E_\mathrm{i})} + \sum_{i}\frac{\left[I_\mathrm{H}(E_\mathrm{i})-I_\mathrm{M}(E_\mathrm{i})\right]^2}{\sigma_\mathrm{H}^2(E_\mathrm{i})},
\end{equation}
where the summations are over the measured energy points $E_\mathrm{i}$. $(I_\mathrm{D},\sigma_\mathrm{D})$, and $(I_\mathrm{H},\sigma_\mathrm{H})$ are the measured electron plus positron flux and the associated errors from the DAMPE and the H.E.S.S. experiments, respectively, at their respective $E_\mathrm{i}$'s. $I_\mathrm{M}(E_\mathrm{i})=I_\mathrm{e^-}(E_\mathrm{i})+I_\mathrm{e^+}(E_\mathrm{i})$. As such, $I_\mathrm{e^-}(E_\mathrm{i})=(v/4\pi)N(E_\mathrm{i})$ is the electron flux predicted by the model (Eqs. \ref{eq-sol2} or \ref{eq-sol3}), and $I_\mathrm{e^+}(E_\mathrm{i})$ is the observed positron flux from the AMS-02 experiment, which can be described as \citep{AMS2019b}
\begin{align}
\label{eq-pos}
I_{e^+}(E_\mathrm{i})=\frac{E_\mathrm{i}^2}{\hat{E}^2}\left[K_1\left(\frac{\hat{E}}{E_1}\right)^{\gamma_1}+K_2\left(\frac{\hat{E}}{E_2}\right)^{\gamma_2} \exp\left(-\frac{\hat{E}}{E_\mathrm{c}}\right)\right],
\end{align}
where the first term represents contribution of the diffuse positron flux produced from the collision of CR nuclei with interstellar matter, and the second term represents an additional contribution of positrons, which could be originated from dark matter annihilation or astrophysical sources such as pulsars in the Galaxy. $\hat{E}=E_\mathrm{i}+\phi$, with $\phi=1.1$~GeV, is the solar modulation parameter, $K_1=6.51\times 10^{-2}$~$(\mathrm{m^2\,sr\,s\,GeV})^{-1}$, $K_2=6.80\times 10^{-5}$~$(\mathrm{m^2\,sr\,s\,GeV})^{-1}$, $E_1=7.0$~GeV, $E_2=60$~GeV, $\gamma_1=-4.07$, $\gamma_2=-2.58$, and $E_\mathrm{c}\approx 813$~GeV. In writing Eq. (\ref{eq-chi2}), we assumed that the data points are uncorrelated.

\begin{figure*}
%\vspace{-0.3cm}
\centering
\includegraphics*[width=\columnwidth,angle=0,clip]{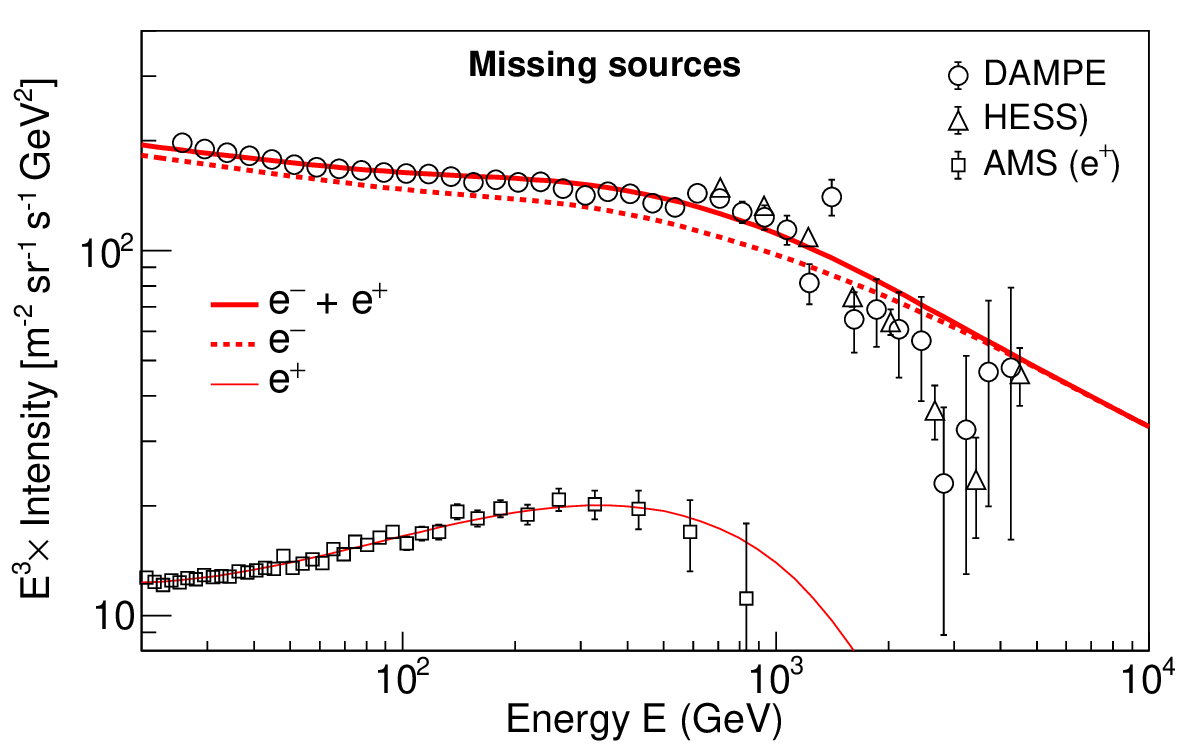}
\includegraphics*[width=\columnwidth,angle=0,clip]{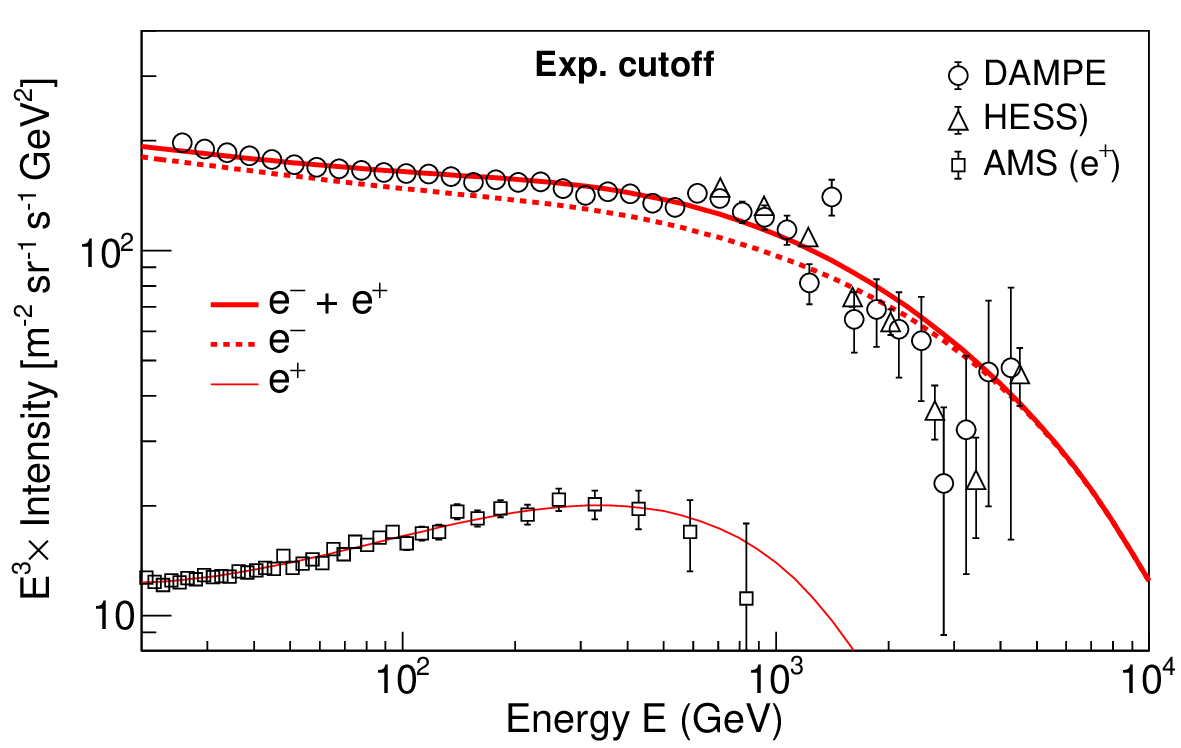}\\
\includegraphics*[width=\columnwidth,angle=0,clip]{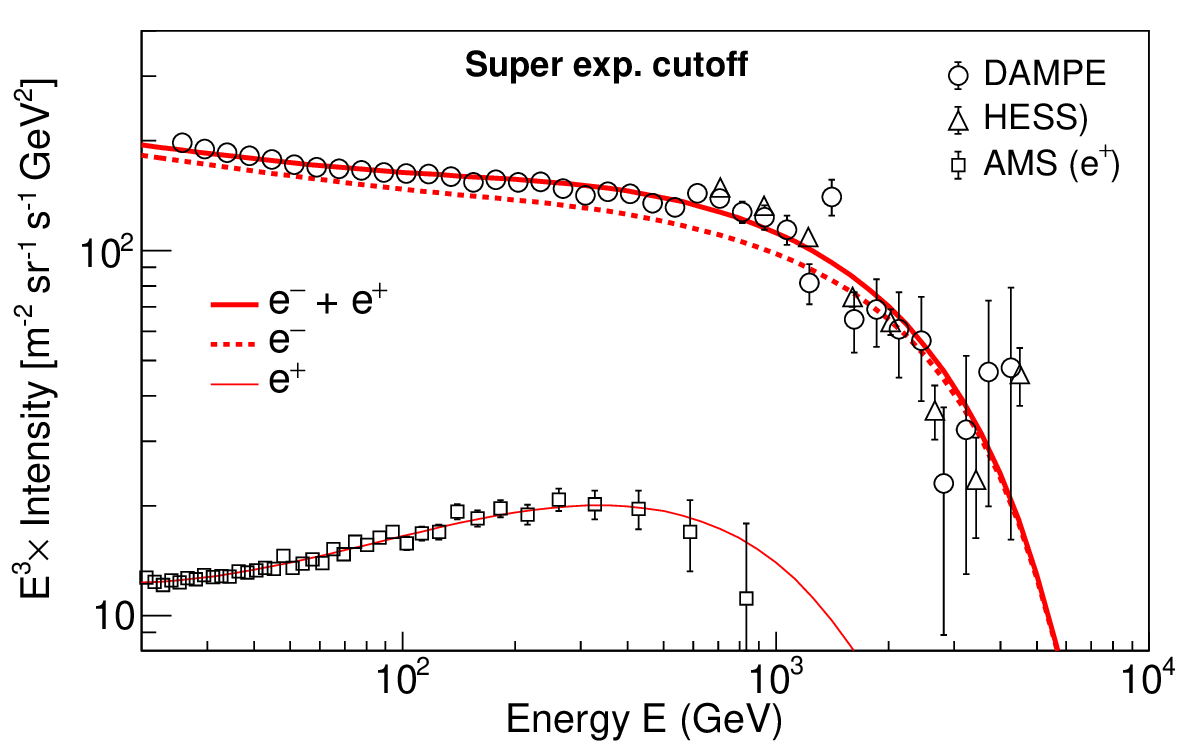}
\includegraphics*[width=\columnwidth,angle=0,clip]{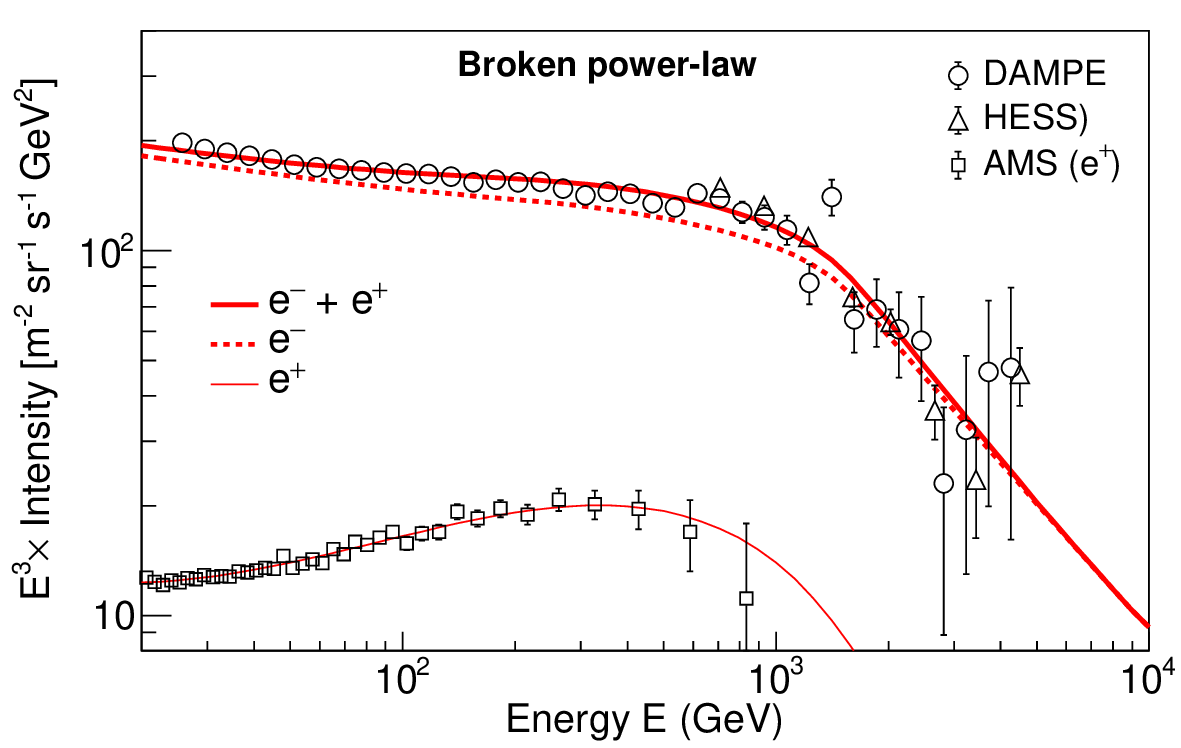}
\caption{\label {fig-1} Fit to the CR electron plus positron data from the DAMPE \citep{DAMPE2017} and the H.E.S.S. \citep{HESS2008} experiments for the $(20-3500)$~GeV fit range for different scenarios of the spectral beak: missing sources (top left), exponential cutoff (top right), super-exponential cutoff (bottom left), and broken power-law (bottom right). Line representation: best-fit electron plus positron spectrum (thick solid line), electron spectrum (dashed line) and positron spectrum (thin solid line). The positron spectrum and its data are from AMS-02 measurements  \citep{AMS2019b}. The fits exclude the three highest energy data points. The best-fit values of the model parameters are listed in Table \ref{fit-parameters}.}
%\vspace{0.3cm}
\end{figure*}

In the next step, $(\Gamma,r_0)$ or $(\Gamma,E_0)$ were fixed to the values that maximize $\mathcal{L}$, and we scanned over $f$ in finer steps of $\Delta f=0.001$ to determine its best-fit value that gives the minimum value of $\chi^2$. Table \ref{fit-parameter-range} lists the model parameters and their ranges used in the fit. The fit was performed for the energy range $(20-3500)$~GeV. This range excludes the three highest energy data points of the measured spectrum shown in Fig. 1. Although measurement uncertainties are large at these energies, the rising trend in the flux might indicate the presence of one or several strong nearby sources, which we did not include in this study \citep{Kobayashi2004}. Considering SNRs located within $1$~kpc of Earth and using a reasonable form of energy-dependent CR escape from the SNRs, \cite{Thoudam2012} showed that nearby sources, in particular the Vela remnant, can be responsible for the steep rise in the observed spectrum at the highest energies. We note that for the energy range considered in this work, the effects of processes such as solar modulation, re-acceleration by interstellar magnetic turbulence or by encounters with old SNRs, convection by the Galactic wind, and energy losses due to ionization and bremsstrahlung on the electron spectrum at Earth are expected to be negligible. These processes are important mostly at energies below $\sim\,20$~GeV.

In order to check the effect of the selected energy range on the fit results, we also performed a fit for another energy range of $(100-3500)$~GeV. The ranges of the model parameters for this fit range are also listed in Table \ref{fit-parameter-range}.

\begin{table*}
\centering
\caption{\label{table-parameters} Best-fit values of the model parameters $(f,\Gamma,E_0, r_0)$ and the $\chi^2$ values of the fits.}
\label{fit-parameters}
\resizebox{\textwidth}{!}
{
\begin{tabular}{cccccc|ccccc}
\hline
\hline
&\multicolumn{5}{c}{Fit range: $(20-3500)$~GeV}  & \multicolumn{5}{c}{Fit range: $(100-3500)$~GeV}\\
\cline{2-6}\cline{7-11}
&&&&&&&&&&\\
&$f$                                    &$\Gamma$       &$E_0$          &$r_0$  &$\chi^2$               &$f$                                    &$\Gamma$       &$E_0$  & $r_0$   &$\chi^2$\\
&($\times 10^{49}$~ergs)        &                       &(GeV)          &(kpc)  &($ndf=43$)     &($\times 10^{49}$~ergs)  &                       &(GeV)  & (kpc)         &($ndf=33$)\\
    
\hline
&&&&&&&&&&\\
Missing sources         &       $0.272$ &$2.23$ & $-$   & $0.85$        & $125$ & $0.378$ &       $2.08$ & $-$            & $1.3$         & $111$\\
Exponential cutoff             &       $0.154$ &$2.37$         & $6762$ & $-$   & $114$ & $0.116$       &       $2.28$ & $4250$         & $-$   & $89$\\
Super-exponential cutoff       &       $0.159$ &$2.39$         & $4588$ & $-$   & $87$ & $0.125$        &       $2.32$ & $3628$         & $-$   & $67$\\
Broken power-law                &       $0.159$ &$2.39$         & $1576$ & $-$   & $69$ & $0.142$        &       $2.36$ & $1482$         & $-$   & $57$\\
\hline
\vspace{0.2cm}
\end{tabular}
}
\tablefoot{The values listed are for the two different fit ranges: $(20-3500)$~GeV and $(100-3500)$~GeV. The $\chi^2$ calculation includes both the DAMPE and the H.E.S.S. data.}
\end{table*}

\begin{figure*}
%\vspace{-0.3cm}
\centering
\includegraphics*[width=\columnwidth,angle=0,clip]{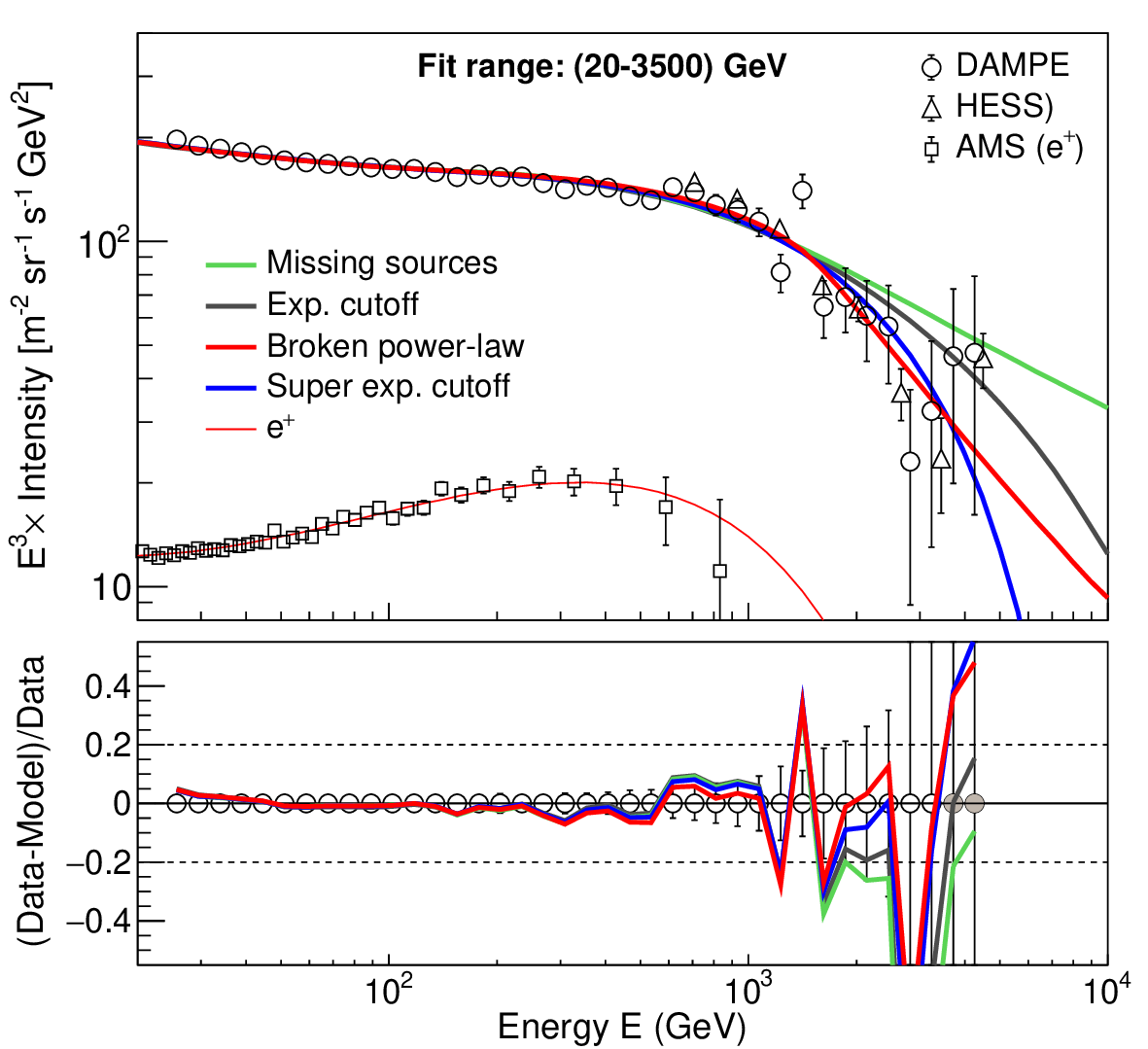}
\includegraphics*[width=\columnwidth,angle=0,clip]{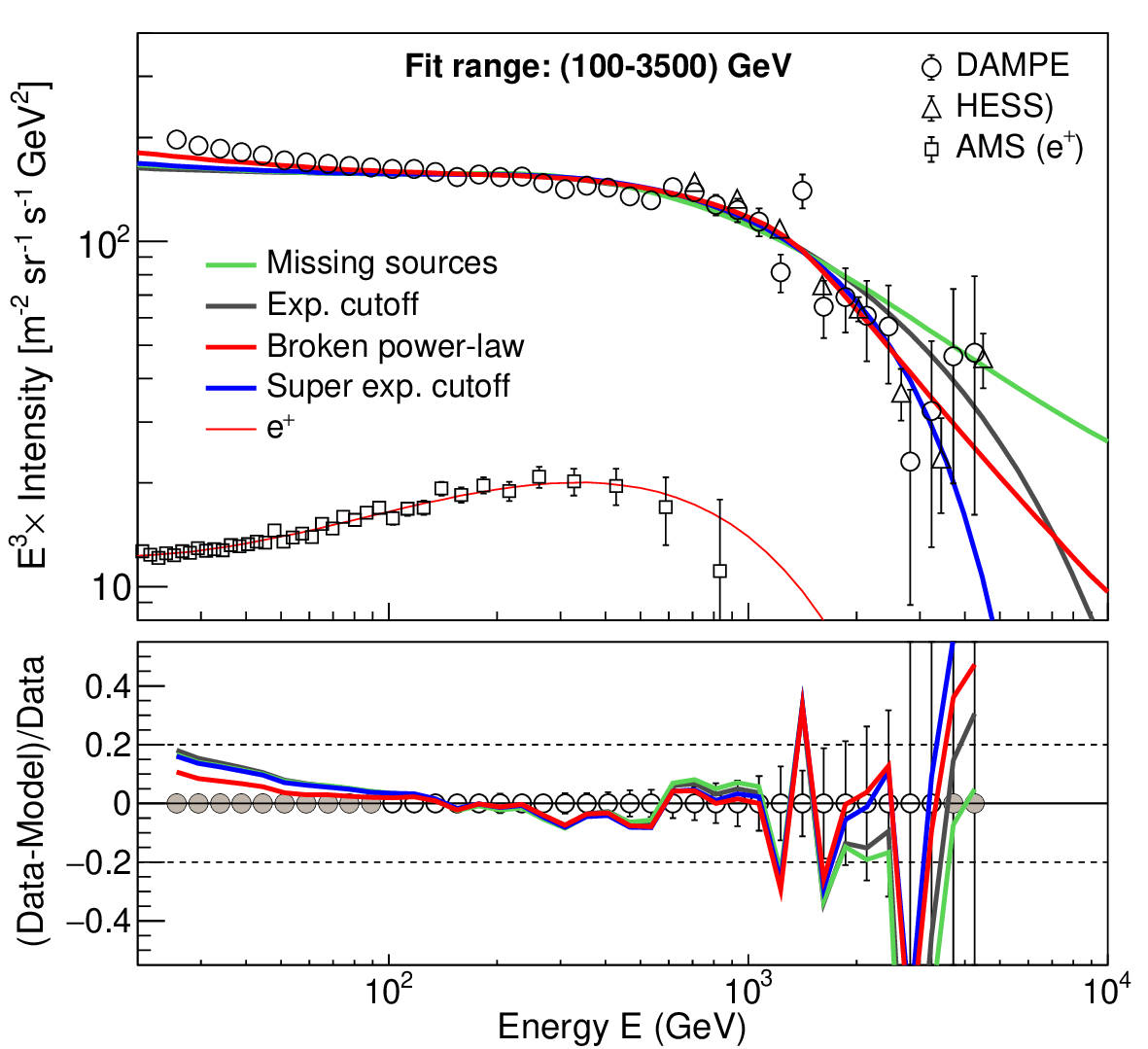}
\caption{\label {fig-2} Comparison of the best-fit electron plus positron spectra of the different scenarios for the two different fit ranges: $(20-3500)$~GeV (left) and $(100-3500)$~GeV (right). Line representation: missing sources (green line), exponential cutoff (black line), broken power-law (thick red line), and super-exponential cutoff (blue line). For the $(20-3500)$~GeV fit, the best-fit electron plus positron spectra are the same as shown in Fig. 1. The top plots show the spectral fits, and the bottom plots show the fit residuals. Residuals are shown only with respect to the DAMPE data, but the $\chi^2$ calculation  includes both the DAMPE and the H.E.S.S. data. Filled circles in the residual plots represent data points not included in the fit. The dashed lines represent $\pm20\%$ residual levels, shown only for reference. The best-fit values of the model parameters and the $\chi^2$ values are listed in Table \ref{fit-parameters}. All of the data shown and the positron spectrum (thin red line) are the same as in Fig. 1.}
%\vspace{0.3cm}
\end{figure*}

\section{Results}
\label{sec-result}
The fit results for the $(20-3500)$~GeV range are shown in Fig. 1 for the different scenarios of the spectral break where the best-fit electron plus positron spectrum (thick solid line), the electron component (dashed line) and the positron spectrum (thin solid line) are shown along with the  measurements from the DAMPE and the H.E.S.S experiments. The positron spectrum in Fig. 1, given as Eq. \ref{eq-pos}, is a basic fit to the observed data from AMS-02 \citep{AMS2019b}, which is also shown in the figure.

All of the scenarios produce a reasonably good fit to the data up to energies close to $\sim\,1$~TeV, whereas above this energy, the broken power-law and the super-exponential cutoff scenarios show significantly better fits. The missing source and exponential cutoff scenarios did not produce spectra which are steep enough to explain the observed steepening above $1$~TeV. Figure 1 also demonstrates that the positron fluxes do not account for the observed spectral break. The break is determined by the electron component.

To compare the quality of the fits between the different scenarios for the $(20-3500)$~GeV fit range, we plotted the best-fit electron plus positron spectra of the different scenarios from Fig. 1 together into a single plot in Figure 2 (top-left panel). It can be seen that all the scenarios predict an almost similar spectrum at lower energies, but above $\sim\,1.5$~TeV, their predictions start to deviate from each other, with the broken power-law (thick red line) and the super-exponential cutoff (blue line) scenarios explaining the data better than the missing sources (green line) and the exponential cutoff (black line) scenarios. The positron spectrum and its data shown in Fig. 2 are the same as in Fig. 1. The residuals of the spectral fits are shown in the bottom-left panel, calculated only with respect to the DAMPE data, although both DAMPE and H.E.S.S data were used in the fits. The filled circles in the residual plots show the data points that were not included in the fit, and the dashed lines show the $\pm20\%$ residual levels to guide the eye. The right panels of Fig. 2 show similar plots for the $(100-3500)$~GeV fit range. Above $100$~GeV, the individual spectra and the relative differences between them are similar to those observed in the $(20-3500)$~GeV fit range.

Table \ref{fit-parameters} gives the best-fit values of the different parameters, that is, $(f,\Gamma,r_0)$ for the missing sources scenario and $(f,\Gamma,E_0)$ for the other cases, along with the $\chi^2$ values of the fits. Based on the $\chi^2$ values, the broken power-law is found to give the best fit of all the scenarios for both the fit ranges. For the $(20-3500)$~GeV fit, it gives the smallest $\chi^2$ value of $69$ for a number of degrees of freedom of $ndf=43$. The next best case, which is the super-exponential cutoff, gives $\chi^2=87$ for the same $ndf$. The exponential cutoff and the missing sources give $\chi^2=114$ and $125$, respectively. For the $100-3500$~GeV fit, we find similar results, where $\chi^2=57$ for the broken power-law compared to $\chi^2=67$ for the super-exponential cutoff, $\chi^2=89$ for the exponential cutoff, and $\chi^2=111$ for the missing sources scenario for $ndf=33$. For both of the fit ranges, we find that the overall fit quality decreases in the following order of the scenarios: broken power-law, super-exponential cutoff, exponential cutoff, and missing sources. This demonstrates the robustness of the fit with respect to the chosen energy range. 

For the $(20-3500)$~GeV fit, which is the main result of this work, we find that the broken power-law scenario gives $\Gamma=2.39$, $E_0=1576$~GeV, and $f=0.159\times 10^{49}$~ergs, whereas the $(100-3500)$~GeV fit gives a flatter source index of $\Gamma=2.36$ together with smaller $E_0$ and $f$ values of $E_0=1482$~GeV and $f=0.142\times 10^{49}$~ergs. This is because the $(20-3500)$~GeV fit has to accommodate for the observed steeper spectrum below $\sim\,100$~GeV, which then requires larger values of $E_0$ and $f$ to explain both the observed spectral break and the flux level at the same time. For the exponential and the super-exponential cutoffs, we find the best-fit values for the $(20-3500)$~GeV fit to be ($\Gamma=2.37$, $E_0=6762$~GeV, $f=0.154\times 10^{49}$~ergs) and ($\Gamma=2.39$, $E_0=4588$~GeV,  $f=0.159\times 10^{49}$~ergs). These values are also larger than the corresponding values obtained for the $(100-3500)$~GeV fit. The missing sources scenario gives $\Gamma=2.23$, $r_0=0.85$~kpc, and $f=0.272\times 10^{49}$~ergs for the $(20-3500)$~GeV fit, compared to the values of $\Gamma=2.08$, $r_0=1.3$~kpc, and $f=0.378\times 10^{49}$~ergs for the $(100-3500)$~GeV range. This can be explained as: a flatter $\Gamma$ will require a larger $r_0$ value (which corresponds to a lower cutoff energy) to compensate for the increase in the flux at higher energies, due to the flatter spectrum.

We note that our best-fit value for the source index value (for instance, $\Gamma=2.39$ for the broken power-law scenario) is larger than the typical value of $2.0$ predicted by the standard theory of DSA for strong shocks. However, this larger value of $\Gamma$ is typical for CR propagation studies (see e.g., \citealp{Fermi2012}). A similar value of $\Gamma=2.39$ for CR electrons was also found in a recent study presented in \cite{Evoli2020}. It is possible that this type of steep index is produced by CR modified shocks inside SNRs, as discussed recently in \cite{Caprioli2020}, or by the presence of an efficient magnetic field amplification operating until the late stages of SNR evolution, when the shock already becomes weak \citep{Diesing2019, Cristofari2021}.

\section{Discussions and conclusions}
\label{sec-discuss}
We have explored different scenarios related to CR source properties for the origin of the observed break in the CR electron spectrum at $\sim\,1$~TeV, and find that the scenario with the broken power-law source spectrum best describes the data with the best-fit values of $\Gamma=2.39$, $E_0=1576$~GeV, and $f=0.159\times 10^{49}$~ergs. While the widely adopted exponential cutoff scenario as well as the scenario of the missing nearby sources are statistically disfavored, we find that the super-exponential cutoff also explains the data reasonably well, although it is statistically less favorable than the broken power-law case. Moreover, multi-wavelength observations of SNRs provide a strong indication favoring the broken power-law scenario for the TeV break, as discussed below.

DSA applied to supernova shocks in a regular ISM predicts an exponential or super-exponential cutoffs in the electron spectrum depending on the limiting condition for the maximum energy, as discussed in Sect. \ref{sec-scenarios}, but the cutoffs are expected to be at $\sim\,(30-100)$~TeV \citep{Reynolds2008}, which is much higher than the position of the observed break in the CR electron spectrum. Moreover, combined spectral fitting of radio and X-ray data of several young SNRs have also found a similar range of very high energy cutoffs for electrons \citep{Reynolds1999}, thereby disfavoring the exponential or super-exponential cutoffs in the source spectrum as a plausible explanation for the observed TeV break.

On the other hand, DSA simulations using standard SNR parameters have predicted a break in the power-law spectrum of electrons at GeV$-$TeV range (where the position of the break depends on the age of the remnant) arising from the radiative losses of electrons during their confinement in the downstream region \citep{Kang2011}. The break energy which is given as the condition, $t_\mathrm{loss}=t_\mathrm{age}$, follows $E_\mathrm{b}\approx1/(\alpha t_\mathrm{age})$ for a continuous injection of electrons into the downstream (see e.g., \citealp{Thoudam2011}). For synchrotron losses, $\alpha=2.5\times 10^{-18}(B/\mu\mathrm{G})^2$~GeV$^{-1}$~s$^{-1}$, indicating that our best-fit break energy of $E_0=1576$~GeV~$\approx1.6$~TeV corresponds to an SNR age of $t_\mathrm{age}\approx3.3\times10^4$~yr for a conservative magnetic field strength of $B=15\,\mu$G in the downstream. This age is close to the onset of the radiative phase of an SNR, $t_\mathrm{r}=2.7\times10^4(E_\mathrm{SN}/10^{51}\mathrm{ergs})^{0.24}(n_\mathrm{H}/\mathrm{cm^3})^{-0.52}$~yr $\approx2.7\times10^4$~yr, estimated for a standard supernova explosion energy of $E_\mathrm{SN}=10^{51}$~ergs and ISM density of $n_\mathrm{H}=1$ H cm$^{-3}$ \citep{Sturner1997}. It is generally considered that as an SNR enters the radiative phase, the shock slows down significantly and all of the particles confined downstream are released into the ISM. It is worth keeping in mind that our estimate of $t_\mathrm{r}$ does not take into account the variety of interstellar environments in which supernova explosions can occur in the Galaxy. While thermonuclear or Type Ia supernova explosions generally occur in the regular ISM, core-collapse or Type II supernova explosions can occur in a more complex environment, which can be a low-density cavity formed by powerful winds originated from their massive progenitor stars (O-type stars) or a high-density environment of molecular clouds if the progenitors are of lower mass, such as B-type stars \citep{Weaver1977, Chevalier1999}. In addition, most supernovae are expected to be found within star clusters where the environment can be even more complex due to the interacting stellar winds \citep{Longmore2014}. For a typical range of $n_\mathrm{H}=(0.1-10)$ H cm$^{-3}$, we estimate the start of the radiative phase to be in the range of $t_\mathrm{r}\approx(8\times10^3-9\times10^4)$~yr. The SNR age of $t_\mathrm{age}\approx3.3\times10^4$~yr that we obtained corresponding to our best-fit break energy of $E_0\sim1.6$~TeV falls well within this range.

Strong observational evidence in favor of the broken power-law case comes from the extensive H.E.S.S. measurement of SNR $\mathrm{RX\,J}1713.7-3946$, which is one of the most well-studied SNRs across different wavelengths \citep{HESS2018}. Simultaneous fitting of high-quality X-ray and GeV$-$TeV gamma-ray data shows a clear preference for a broken power-law particle spectrum which, for the leptonic origin, requires a power-law break at $E_\mathrm{b}=2.5$~TeV with an index $\Gamma=1.78$  $(2.93)$ below (above) the break and an exponential cutoff of $E_\mathrm{c}\sim\,88$~TeV for a magnetic field strength of $B\sim\,14\,\mu$G. The $\Gamma$ value below the break obtained by the H.E.S.S measurement is smaller than our best-fit value of $\Gamma=2.39$, but the spectral difference of $\Delta\Gamma\sim\,1.15$ found with H.E.S.S. below and above the break is quite close to the $\Delta\Gamma\sim\,1.0$ predicted by the power-law scenario in the present study, which is also the observed difference in the CR electron spectrum. In another study, a fit to the multi-wavelength data using a broken power-law electron spectrum with a super-exponential cutoff gives a slightly different result: $\Gamma=2.0$ below the break, $E_\mathrm{b}=3.5$~TeV, $E_\mathrm{c}=80$~TeV, and $B=16\,\mu$G \citep{Zang2019}. Although the exact origin (leptonic or hadronic) of the non-thermal emission from $\mathrm{RX\,J}1713.7-3946$ is still under debate, we note that the hadronic scenario requires certain environmental conditions for the SNR to generate a broken power-law source spectrum, for instance, a supernova explosion in a molecular cloud that has already been swept away by a strong wind from the progenitor star \citep{Gabici2014}.  Such an environment may not be readily available for the majority of the SNRs in the Galaxy. On the other hand, in the leptonic case, a power-law spectral break is naturally expected, as discussed above in Sect. \ref{bpl}.

Additional evidence for the broken power-law case can be found in the multi-wavelength studies of several other SNRs. \cite{Zang2019} studied a sample of seven middle-aged (several thousand years old) shell-type SNRs and finds that the majority of the SNRs in their sample show an electron spectrum with a power-law break at $E_\mathrm{b}\sim\,(1.0-3.5)$~TeV and $\Gamma\sim\,1.9-2.2$ below the break, along with a high-energy super-exponential cutoff in the range of $\sim\,(30-80)$~TeV and a magnetic field in the range of $B\sim\,(9-26)\,\mu$G. 

\cite{Zeng2019} considered a larger sample of $35$ middle-aged and moderately old SNRs, and also finds a particle distribution that follows a broken power-law spectrum. They find the break energy to decrease with the SNR age in the range of $\sim\,100\,\mathrm{TeV}-1\,\mathrm{GeV}$ over $\sim\,(10^3-6\times10^4)$~yr age range with a spectral index $\Gamma\sim\,1.2-2.3$ below the break energy. However, they considered a combined population of electrons and protons to fit the multi-wavelength data, and assumed the two population to have the same spectral index and break energy, which may not necessarily be true in all cases \citep{Diesing2019,  Morlino2021}. These reservations make it difficult infer information from their results that can be directly compared with our findings in this study, where our focus was purely on the electron spectrum.

Overall, the observation of a spectral break with $\Delta\Gamma\sim\,1.0$ in the electron spectra of several SNRs, such as $\mathrm{RX\,J}1713.7-3946$ and others as presented in \cite{Zang2019}, and our independent finding of a similar break and similar $\Delta\Gamma$ in the electron source spectrum from the study of CR propagation in the Galaxy, provide a strong indication of a direct link between the observed CR electrons and their sources. However, the $\Gamma$ values of $\sim\,(1.9-2.2)$ inferred from the SNRs are found to be relatively smaller than our best-fit value of $2.39$ obtained from the CR electron data. The fact that the $\Gamma$ values from the SNRs are quite close to the index predicted by the DSA process indicates that the electrons confined inside the remnants mostly preserve the shape of the spectrum produced at the shocks, except at energies above the break where the spectrum steepens by $\Delta\Gamma\sim\,1.0$ due to radiative cooling. However, what is necessary for the study of CR propagation is the electron spectrum released by SNRs into the ISM. This spectrum is poorly known, and it may be different from the spectrum produced at the shocks. Our findings strongly indicate that a modification (steepening) of the electron spectrum must be happening while the electrons escape through the source region. Recent studies have shown that CR protons escaping from their sources can generate magnetic fluctuations around the source region due to both resonant and non-resonant streaming instabilities \citep{Schroer2021, Schroer2022}. These fluctuations lead to the formation of low-diffusivity CR inflated bubbles surrounding the sources where CRs spend a considerable amount of time before finally being released into the ISM. For electrons, while they diffuse through these bubbles, their spectrum can be changed due to radiative losses. This may explain the observed difference in the spectral index of electrons obtained from SNRs and that required by the Galactic propagation study. A similar argument has also been raised to explain the difference in the source index of protons and electrons, as revealed by CR propagation studies considering that protons and electrons can suffer different spectral changes during their propagation through these bubbles \citep{Schroer2021, Cristofari2021}. The presence of such bubbles or cocoons around potential CR sources has also been proposed to predict the additional contribution of secondary CRs from the sources leading to a hardening in the secondary-to-primary ratios at TeV energies \citep{Cowsik2010}.

Based on the best-fit results obtained in this work using the observed CR electron spectrum and the evidence reported from the multi-wavelength studies of SNRs, we conclude that if SNRs are the main sources of CRs in the Galaxy, the power-law break in the measured CR electron plus positron spectrum at $\sim\,1$~TeV is most likely an imprint of the radiative cooling break present in the electron source spectrum. Our findings further show that in order to reconcile the difference between the spectral index of electrons obtained from the study of SNRs and CR propagation, CR electrons must undergo spectral modifications while escaping the source region. A detailed investigation of this topic can help further strengthen the link between SNRs and CRs in the Galaxy.

\begin{acknowledgements} Fundings from the ADEK grant (AARE19-224), and the Khalifa University ESIG-2023-008 and RIG-S-2023-070 grants are  acknowledged.
\end{acknowledgements}

\end{document}